# The maximum storage capacity of open-loop written RRAM is around 4 bits

Yongxiang Li [1], Shiqing Wang [1], Zhong Sun*[1,2]

[1] School of Integrated Circuits, Institute for Artificial Intelligence, Peking University
[2] Beijing Advanced Innovation Center for Integrated Circuits

* Email: yongxiang.li@stu.pku.edu.cn, zhong.sun@pku.edu.cn

*Abstract*—There have been a plethora of research on multi-level memory devices, where the resistive random-access memory (RRAM) is a prominent example. Although it is easy to write an RRAM device into multiple (even quasi-continuous) states, it suffers from the inherent variations that should limit the storage capacity, especially in the open-loop writing scenario. There have been many experimental results in this regard, however, it lacks a comprehensive analysis of the valid multi-bit storage capability, especially in theoretical terms. The absence of such an insight usually results in misleading conclusions that either exaggerate or underestimate the storage capacity of RRAM devices. Here, by the concept of information theory, we present a model for evaluating the storage capacity of open-loop written RRAM. Based on the experimental results in the literature and the test results of our own devices, we have carefully examined the effects of number of pre-defined levels, conductance variation, and conductance range, on the storage capacity. The analysis leads to a conclusion that the maximum capacity of RRAM devices is around 4 bits.

## I. Introduction

With the rapid development of technologies such as internet of things, artificial intelligence, and advanced wireless communications, the amount of data generated have been increasing exponentially [1]. It has caused a strong demand for high-density information storage, which, in turn, flourishes the study on emerging memory technologies, such as RRAM, and phase-change memory (PCM), *etc.* [2]. They combine the nonvolatility property and other high-performance indicators [3]. In particular, the underlying physical mechanisms allow these devices to be programmed in an analog manner, through continuous modulation of the internal state variables [4]. Such a capability is also highly desired in emerging computing paradigms, such as in-memory computing and analog matrix computing [5, 6].

The multi-level RRAM device may be programmed by using various schemes, such as those based on the modulation of compliance current, magnitude or width of gate or electrode voltage, or number of voltage pulses. Ideally, the device conductance may be quasi-continuously modulated, resulting in several tens to hundreds of discrete levels [7, 8]. However, in the open-loop writing with no repeated corrections, such results are prone to be overestimated, due to the inherent device variations. On the other hand, the test results of large-volume devices often give relatively conservative estimations (~3 bits), by using programming strategies such as incremental step pulse and incremental gate voltage [9-13]. Such results are purely empirical, lack of sufficient theoretical guidance for continuous optimization.

The storage capacity of RRAM devices may indeed be enhanced by using the strict while tedious closed-loop verify scheme [7, 14], where each target level with a pre-defined conductance window can be achieved through tens of set or reset operations, resulting in well separated states with no overlap. However, multiple cycles of verification cause high costs of programing latency and energy dissipation, and repeated switching transitions are also harmful to the endurance performance of RRAM devices.

In order to explore the achievable maximum storage capacity of RRAM devices with open-loop writing (which in turn may be used as an optimization guideline), in this work we have studied this issue with the concept of information theory. A comprehensive analysis has been conducted to reveal the impact of number of levels, conductance variation and conductance range on the storage capacity. The results suggest that, under the consideration of common conditions of RRAM devices, particularly the obstacle set by the conductance variation, the maximum storage capacity of open-loop written RRAM is around 4 bits.

## II. Storage Capacity Model

### A. Definition of Storage Capacity

Given that the RRAM device conductance can be determined by external parameters, we consider the open-loop writing of RRAM devices as an analog coding problem. As shown in Fig. 1, the external electrical input is encoded as the conductance output stored in the RRAM device through an operation with inherent noise that finally present an output distribution [15]. Each level features a specific mean conductance value $G$ and a standard deviation $\sigma_G$. This probabilistic correspondence between input and output can be described by the concept of information theory, obtaining the storage capacity of the open-loop writing of RRAM is,

$$C = -\sum_V P(V) \log_2 P(V) + \sum_{V/G} P(V,G) \log_2 P(V|G)$$

**Ideal Capacity**    **Coupling term** (<0)

Fig. 1. Storage capacity model and calculation equation.

$$C = -\sum_V P(V)\log_2 P(V) + \sum_{V,G} P(V,G)\log_2 P(V|G), \quad (1)$$

where $P(V)$ is the probability of taking value of $V$ among all input voltages, $P(G)$ is the probability of $G$ among all output conductances, $P(V|G)$ is the conditional probability of the input voltage $V$ on the output conductance $G$, and $P(V,G) = P(G)P(V|G)$. Eq. 1 consists of two terms, where the first term indicate the ideal situation with no noise distortion. The second term represents the noise coupling relationship between the voltage input and the conductance output, which is always less than 0. Intuitively, it is due to the overlaps between neighboring levels that causes the loss of effective bits of information. It should be stressed that although the input is considered as voltage magnitude in this model, it could also be voltage width, pulse number, or compliance current [14, 16].

### B. Survey of Experimental Results and RRAM Model

To evaluate the storage capacity of open-loop written RRAM devices, we have conducted a survey of the conductance ranges and variations of multi-level RRAM disclosed in the literature, which are based on frequently-used metal oxides as the resistive switching layer, such as $HfO_2$, $TiO_2$, and $Al_2O_3$ [9-13, 16-17]. Fig. 2a shows that although the relationship between $\sigma_G$ and $G$ is noisy, there is a trend that $\sigma_G$ is independent on $G$. The conductance range $G_{min} \sim G_{max}$ integrating the results of the RRAM devices is [1~250] μS, and the conductance variation $\sigma_G$ fluctuates in the range of [3~20] μS.

Based on the observation of constant $\sigma_G$, we have established a simple but reasonable multi-level RRAM model, as shown in Fig. 2b. Different conductance ranges and $\sigma_G$'s will be considered, resulting in different degrees of inter-level overlaps. All conductance levels are evenly distributed in a given conductance range, following normal distributions with the same $\sigma_G$ for each level. The mean values of every two neighboring levels are separated by a conductance difference $\Delta_G$. For some conductance levels at the edge of range, their distributions may overflow beyond the available range, the distribution probabilities are therefore truncated and normalized. To alleviate this issue, we have also reserved a certain amount of space at the edge, that is, the highest and the lowest levels are shifted from $G_{min}$ and $G_{max}$ to the left and right by $\Delta_G/2$, respectively.

### C. Storage Capacity Analysis

Based on the multi-level RRAM model, the equivalent storage capacity $C$ of RRAM device is calculated by using Eq. 1. First, the full range of [1~250] μS is considered, and different numbers of discrete levels are assumed. For each given $\sigma_G$, the evolvement of $C$ increasing with the presumed $L$ is calculated, as shown in Fig. 3. In all cases, the curve increases rapidly when $L$ is small, where the first term in Eq. 1 dominates, and the intervals are sufficiently large to prevent the inter-level overlaps. When $L$ exceeds a certain number ($L_T$), the curve quickly saturates at the final storage capacity $C_{max}$. Because of the growing overlap between conductance levels, the storage capacity is suppressed by the second term in Eq. 1.

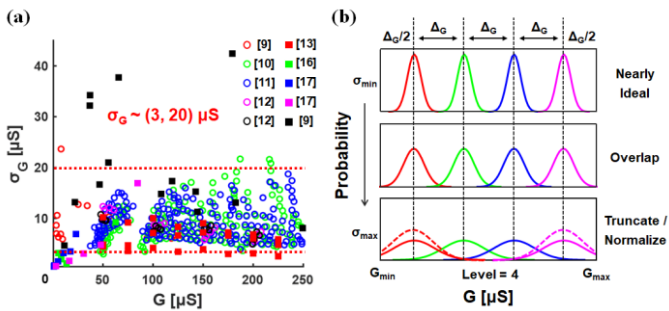

Fig. 2. (a) The $\sigma_G$-$G$ correspondence data in Refs. [9-13, 16-17], which show roughly a constant behavior. (b) Multi-level RRAM model with identical conductance variation for each level.

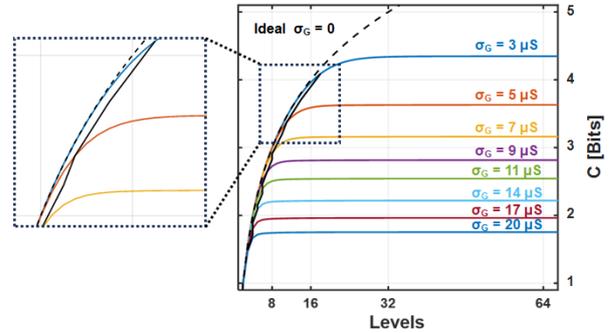

Fig. 3. Growth curve of storage capacity against the pre-defined number of discrete conductance levels. The magnified inset highlights the transitions to saturation capacities.

We provide a quantitative analysis of $C_{max}$ and $L_T$. $C_{max}$ is obtained by calculating with a sufficiently large $L$, and $L_T$ is defined as then number when the capacity $C = 0.95C_{max}$, which are marked by the black line in the inset of Fig. 3. Before the transition at $L_T$, the capacity curve under different $\sigma_G$ conditions are all close to the ideal situation (black dotted line), suggesting the rationality of the definition of $L_T$ for identifying the transition place. After $L_T$, the curve increases quickly from $0.95C_{max}$ to $C_{max}$, within only a few extra levels. The values of $C_{max}$ and $L_T$ are plotted in Fig. 4a. the maximum storage capacity is 4.3 bits under the most modest assumption of conductance variations. It also evidences that $C_{max}$ decreases as $\sigma_G$ increases, which shows a nonlinear behavior, namely the decrease of $C_{max}$ is particularly sensitive when $\sigma_G$ is relatively small. This nonlinear behavior is even stronger for $L_T$, suggesting that when $\sigma_G$ is relatively large, it is not helpful to increase the pre-defined number of levels for storage. We have also calculated the parameter $\Delta_G/\sigma_G$ upon the usage of $L_T$, which are almost constant around 4 (Fig. 4b). It means that the coupling between conductance levels starts to occur approximately when the conductance variation $2\sigma_G$ occupy the inter-level space.

In addition to the full range of [1~250] μS, we have conducted calculations for different conductance ranges (Fig.

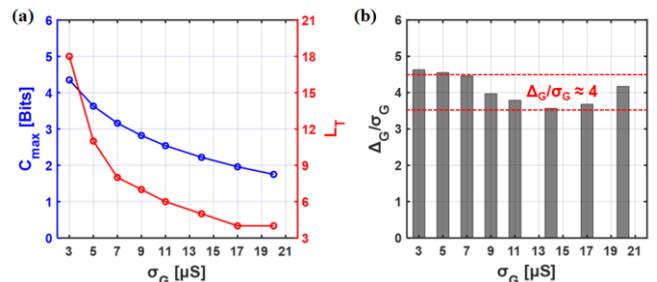

Fig. 4. (a) Dependences of $C_{max}$ and $L_T$ on $\sigma_G$, for the conductance range of [1~250] μS. (b) Relationship between ratio $\Delta_G/\sigma_G$ and $\sigma_G$.

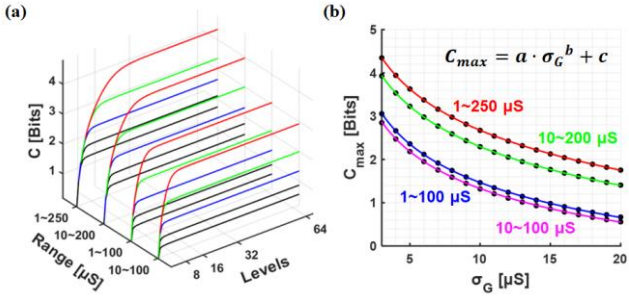

Fig. 5. (a) Growth curve of storage capacity for different conductance ranges. In each case, the considered $\sigma_G$'s from top to bottom are [3, 6, 10, 15, 20] µS. (b) Dependences of $C_{max}$ on $\sigma_G$. For all kinds of conductance range, the relationship between $C_{max}$ and $\sigma_G$ can be fitted by a power function distribution, which suitable for different conductance range.

5a). They all show similar behaviors, demonstrating the universality of the above analysis. For each conductance range, the $C_{max}$ values under different conductance variations $\sigma_G$'s are calculated, resulting in the curves shown in Fig. 5b. As the latter three ranges are narrower, the $C_{max}$ values become smaller, all less than 4 bits. Noticeably, for all the four situations, the nonlinear correspondence between $C_{max}$ and $\sigma_G$ can be perfectly fit by a power function (Fig. 6b), but with different fitting parameters, which again supports the analysis of storage capacity of RRAM devices by the information theory.

### III. EXPERIMENT RESULTS

To support the multi-level RRAM behaviors and the related storage capacity analysis, we have fabricated $HfO_2$-based RRAM devices and carried out extensive measurements of the multi-level characteristics of multiple devices [18]. The multi-level characteristics was studied by externally limiting the compliance current ($I_{cc}$) during the current-voltage (I-V) sweeps (Fig. 6a). For each device, 19 $I_{cc}$'s in the range of [20~200] µA were set for test, resulting in 19 sets of data (30,000 in total). For each $I_{cc}$, the corresponding mean value of conductance and the standard deviation are calculated. The data are summarized in Fig. 6b, which is plotted in the same manner as Fig. 2a. The resulting conductance is in the range [10-200] µS, and the variation of conductance states is in the range of [3~20] µS, which are consistent with the results in the literature, thus supporting the applicability of the above analysis.

Given the correspondence between $\sigma_G$ and $G$ is considerably random (in both the reported conductance standard deviation and the experimental ones), we performed simulations with randomly-assumed $\sigma_G$ for each conductance level. 100 simulations were performed, and the distribution of storage capacity and their average values are shown in Fig. 6c. These data points are all within the envelope of $\sigma_G = 3$ and 20 µS, which corresponds to the best or worst situation of all conductance levels. The average results tell that the real storage capacity of RRAM considering distribution of variations would be less than 3 bits per device. By observing the density of data points, it is concluded that the capacity is sensitive to the deterioration of $\sigma_G$, which can be explained by the exponential decrease of storage capacity in Fig. 5b.

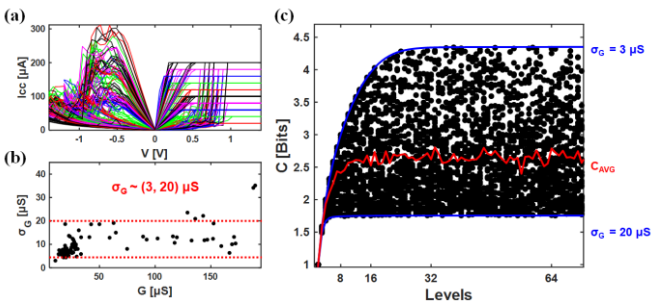

Fig. 6. (a) I-V characteristics of RRAM devices, where the compliance current $I_{CC}$ is used to control the device conductance. The inset illustrates the device structure. (b) $\sigma_G$ of each conductance level in experiment. (c) Comparison of storage capacity curves, under the consideration of uniform distribution (red line) and normal distribution (blue line) for RRAM conductance levels.

### IV. CONCLUSION

In this work, we have established a framework for evaluating the storage capacity of multi-level RRAM devices, by using the concept of information theory. Based on the reported conductance ranges and variations in the literature and those measured in our own RRAM devices, the evaluation results show that the maximum storage capacity of open-loop written RRAM devices is around 4 bits. It provides an effective estimation and an optimization guideline for maximizing the storage capacity of RRAM. We expect that such a framework should also be applicable to other nonvolatile resistive memory devices, to provide a theoretical insight into the capacity limit of storage capacity, thus supporting the delivery of storage-class memory and novel computing paradigms.

### ACKNOWLEDGMENT

This work was supported in part by the National Key R&D Program of China under Grant 2020YFB2206001, in part by NSFC under Grant 92064004 and Grant 61927901, and in part by the 111 Project under Grant B18001.